\documentclass[12pt,dvipdfmx]{article}

\renewcommand{\thefootnote}{\fnsymbol{footnote}}

\usepackage{amsbsy,amssymb,latexsym,amsfonts,amsmath}
\usepackage{mathrsfs}
\usepackage[]{graphicx}
\usepackage{bm}
\usepackage{here}
\usepackage{color}

\numberwithin{equation}{section}
\newcommand{\bel}[1]{\begin{equation}\label{#1}}                     
\newcommand{\bal}[1]{\begin{eqnarray}\label{#1}}                     
\newcommand{\be}{\begin{equation}}
\newcommand{\ee}{\end{equation}}

\begin{document}
	%
	%
\begin{titlepage}
		\begin{flushright}
			\normalsize
			~~~~
			NITEP 92\\
			OCU-PHYS 534\\
			March 22, 2021\\
		\end{flushright}
		
		\vspace{15pt}
		
		\begin{center}
			{\LARGE Theory space of one unitary matrix model   }\\
			\vspace{5pt}
			{\LARGE and its critical behavior associated with }\\
			\vspace{5pt}
			{\LARGE  Argyres-Douglas theory }
		\end{center}
		
		\vspace{23pt}
		
		\begin{center}
			{ H. Itoyama$^{a, b,c}$\footnote{e-mail: itoyama@sci.osaka-cu.ac.jp},
				and Katsuya Yano$^b$\footnote{e-mail: katsuyayano@zy.osaka-cu.ac.jp} }\\
			
			%
			\vspace{10pt}
			%
			
			$^a$\it Nambu Yoichiro Institute of Theoretical and Experimental Physics (NITEP),\\
			Osaka City University\\
			\vspace{5pt}
			
			$^b$\it Department of Mathematics and Physics, Graduate School of Science,\\
			Osaka City University\\
			\vspace{5pt}
			
			$^c$\it Osaka City University Advanced Mathematical Institute (OCAMI)
			
			\vspace{5pt}
			
			3-3-138, Sugimoto, Sumiyoshi-ku, Osaka, 558-8585, Japan \\

		\end{center}
		%
		\vspace{10pt}
		\begin{center}
			Abstract\\
		\end{center}
			The lowest critical point of one unitary matrix model with cosine plus logarithmic potential is known to correspond with the $(A_1, A_3)$ Argyres-Douglas (AD) theory
and its double scaling limit derives the Painlev\'{e} II equation with parameter.
Here, we consider the critical points associated with all cosine potentials and determine the scaling operators, their vevs and their scaling dimensions from perturbed string equations at planar level.
These dimensions agree with those of $(A_1,A_{4k-1})$ AD theory.

		
		\vfill

\end{titlepage}
	
\renewcommand{\thefootnote}{\arabic{footnote}}
\setcounter{footnote}{0}

\section{Introduction}

Matrix models are useful in the analysis of four dimensional  supersymmetric gauge theory.
The instanton partition function \cite{Nek, NY} of the four dimensional $SU(2)$ linear quiver superconformal gauge theories can be identified with the conformal block of regular vertex operators in two dimensional conformal field theory by the AGT correspondence \cite{AGT}.
The integral representation of the conformal block \cite{DF} is the multi-Penner type $\beta$-deformed matrix model whose potential has logarithmic terms.

The simplest example is given by the four dimensional $\mathcal{N}=2$ $SU(2)$ supersymmetric gauge theory with four hypermultiplets.
The instanton partition function corresponds to a four point conformal block by the regular vertex operators whose integral representation is the three-Penner type matrix model \cite{ MMS10,IO5}.
The instanton partition functions for the cases with less than four flavors which are asymptotically free are obtained by taking a degeneration limit of the regular vertex operators \cite{G09, GT}.
As a result, an irregular conformal block is formed.
The emergent matrix model contains rational terms \cite{EM, IOYone}.
A similar limit can be taken for the four dimensional $\mathcal{N}=2$ $SU(2)$ linear quiver superconformal gauge theories and we can obtain corresponding asymptotically free theories.
The potentials of these matrix models contain rational terms with higher powers \cite{EM, NR}.
For reviews, see\cite{Maru, IYoshi, Itoyamagakkai}.

Connection with integrable systems is understood better by considering a discrete Fourier transform of the instanton partition function at $\beta = 1$ \cite{MM17} with regard to the filling fraction, which is the Coulomb moduli\footnote{The generating function of the $q$-deformed matrix model has been considered in \cite{MMZ}.}.
Connection between the tau function of Painlev\'{e} equations and the generating function of the instanton partition function of four dimensional $\mathcal{N}=2$ $SU(2)$ gauge theory with $N_f = 0,\ldots,4$ and that of Argyres-Douglas (AD) type theories \cite{AD} obtained in \cite{APSW, KY} \footnote{References in the higher rank case include \cite{EHIY, IKO19, IKO20}.}  has been pointed out in \cite{GIL12, GIL13, ILT,nag1611,BLMST}.
Some of these cases can be derived by using matrix models \cite{IOYanok1, IOYanok2, IOYpro}.
In \cite{IOYanok1, IOYanok2,MOT}, it was shown that in the case of two flavors can be represented by the unitary matrix model of potential $U + U^\dagger + \log U$ type.
This is a Gross-Witten-Wadia (GWW) model \cite{GW80, wad1212, wad80} with a logarithmic potential.
The string equations, which are a set of difference equations arising from the recursion relations among orthogonal polynomials, have been shown to be  the alternate discrete Painlev\'{e} II equations (alt-dPII)\footnote{It is also called the discrete Painlev\'{e} equation $\mathrm{d}$-$\mathrm{P}((2A_1)^{(1)}/D_6^{(1)})$ in \cite{KNY}. }. 
We have also shown that the partition function of this model is the tau function of alt-dPII equation which is closely related to the Painlev\'{e} ${\rm III}_1$ equation.
Using the partition function of this model, we have constructed the tau function of Painlev\'{e} ${\rm III}_1$ equation.
By taking the double scaling limit, alt-dPII equation turns into  Painlev\'{e} II equation with accessory parameter\footnote{For an alternative approach based on the genus expansion of two-cut Hermitian cubic model, see \cite{GG}.}.

Extension of GWW model plus a logarithmic potential is given by a potential with higher powers of $U$ and those of $U^\dagger$.
It was considered in \cite{BMS}  without a logarithmic potential.
(For the recent development about the phase structures of a generalized GWW model, see \cite{ST}. )
It is natural to identify the generating functional of this extended unitary matrix model with that of $\hat{A}_{2m,2n}$ \cite{CV, BMT}  asymptotically free theory.
In this view point, the multicritical points are AD points of various type.
In fact, in \cite{IOYanok3}, we have taken the double scaling limit at the next to the lowest order critical point and derived the system of differential equations associated with the $(A_1, A_7)$ AD theory.

In this paper we develop a further extension of this type.
We will examine the planar critical behavior at various critical points.
It is known that the critical points of the symmetric unitary matrix model are labeled by an integer $k$ \cite{PS1,PS2}.
The analysis of critical behavior and the construction of the even type scaling operators which perturb these critical points are given there.\footnote{See \cite{Kaz} for the hermitian matrix case.}
We take the viewpoint that the Coulomb branch operators of some AD theory corresponds to these scaling operators.
Note that not all of the Coulomb branch operators are reproduced by \cite{PS1,PS2}.
We will construct scaling operators of odd type and logarithmic type in addition to even type and derive their vevs,
and determine their scaling dimensions from these.
By comparing their dimensions with those of the Coulomb branch operator from AD theory, we give further evidence for the correspondence between the $k$-th multicritical point of unitary matrix model and the $(A_1, A_{4k-1})$ AD point.

The paper is organized as follows.
In section $2$, we first review the method of orthogonal polynomial for a unitary matrix model and explain how planar string equations are given.
Then we derive an explicit form of these equations at the $k$-th multicritical point.
We determine the vevs and their dimensions of scaling operators of various type.
In the last section, we give a dictionary between the scaling operators at the $k$-th multicritical point and the Coulomb branch operators in the $(A_1, A_{4k-1})$ AD theory by comparing their scaling dimensions.

\section{One  unitary matrix model}

The partition function of the one unitary matrix model is defined by
\begin{align}
	Z_U = \frac{1}{N !} \left( \prod_{i=1}^N \oint \frac{d z_i}{2\pi i z_i} \right) \Delta(z) \Delta(z^{-1}) \exp\left( \sum_{i=1}^N W (z_i) \right), \label{partition function}
\end{align}
where $\Delta(z)$ is the Vandermonde determinant $\Delta (z) = \prod_{i < j} (z_i - z_j)$ and $W(z)$ is the potential
\begin{align}
	W (z) = -\frac{1}{2 \underline{g}_s} \left[ \sum_{p=1}^\infty \frac{g^+_p}{p} \left( z^p +\frac{1}{z^p} \right) + \frac{g^-_p}{p} \left( z^p -\frac{1}{z^p} \right)  \right].
\end{align}
When all $g^-_p$ vanish, the model is called symmetric.
We can perform the large $N$ expansion for the free energy of \eqref{partition function}:
\begin{align}
	\mathcal{F} \equiv \log Z_U = \sum_{g=0}^\infty N^{2 - 2g} F_g (\widetilde{S}). \label{large N}
\end{align}
The coefficients in the expansion are the function of the parameters in the potential $\{ g^+_p,\,g^{-}_p \}$ and the t' Hooft coupling $\widetilde{S} \equiv \underline{g}_s N$ which is fixed in the large $N$ limit.

A way to evaluate the free energy is to resort to the method of orthogonal polynomials \cite{PS1,PS2,MP90,ger54, ger60,ger62, sim1}.
For a review in the hermitian case, see \cite{LAG} and other approaches, in particular, based on Virasoro constraints, see \cite{IM, Itoyama, MM90, David}.
Let us introduce the set of monic orthogonal polynomials $\{ p_n(z) , \, \tilde{p}_n(1/z) \}$. 
Their orthogonality condition with regard to the measure is
\begin{align}
	\oint d\mu(z) p_n(z) \tilde{p}_m(1/z) = h_n \delta_{n, m},\qquad d\mu(z) \equiv \frac{dz}{2 \pi i z} e^{W(z)}. \label{ortho cond}
\end{align}
Here, $p_n$$(\tilde{p}_n)$ is the polynomial in $z$$(z^{-1})$ of degree $n$
\begin{align}
	p_n(z) = z^n + \cdots + A_n, \qquad \tilde{p}_n(1/z) = z^{-n} + \cdots + B_n.
\end{align}
We have denoted the constant terms by $A_n \equiv p_n(0),\, B_n \equiv \tilde{p}_n (0)$.
They are related to $h_n$ by
\begin{align}
	\frac{h_n}{h_{n-1}} = 1 - A_n B_n. \label{relhAB}
\end{align}
From the orthogonality condition \eqref{ortho cond}, one can show that these polynomials obey the following recursion relations
\begin{align}
	z p_n(z) =& p_{n+1}(z) - \sum_{k=0}^n \frac{h_n}{h_k} A_{n+1} B_k p_k(z) ,\label{rec1} \\
	z^{-1} \tilde{p}_n(1/z) =& \tilde{p}_{n+1} (1/z) - \sum_{k=0}^n \frac{h_n}{h_k} A_k B_{n+1} \tilde{p}_k(1/z). \label{rec2}
\end{align}
Rewriting the Vandermonde determinant in \eqref{partition function} as
\begin{align}
	\Delta(z) = \det (p_{j-1}(z_i))_{1 \leq i, j \leq N},\qquad \Delta(1/z) = \det (\tilde{p}_{j-1}(1/z_i))_{1 \leq i, j \leq N},
\end{align}
and using the orthogonality, we have
\begin{align}
	Z_U  = \prod_{k=0}^{N-1} h_k = h_0^{N} \prod_{j=1}^{N-1} (1 - A_j B_j)^{N-j} \label{pfortho}.
\end{align}
Then, in terms of the coefficients in orthogonal polynomials, the free energy is written as
\begin{align}
	\mathcal{F} = N \log h_0 + \sum_{j=1}^{N-1} (N - j)\log (1 - A_j B_j). \label{free energy}
\end{align}
In particular, the planar free energy $F_0$ can be evaluated as
\begin{align}
	F_0 \sim \int_0^1 dx (1 - x) \log (1 - A(x)B(x)),\qquad \frac{n}{N} \rightarrow x,\,A_n \rightarrow A(x),\,B_n \rightarrow B(x) \label{planar free}.
\end{align}
Here $\sim$ means  dropping the higher contributions in $1/N$.

\subsection{string equation and $k$-th multicritical points}

It is known that the parameter space of the symmetric unitary matrix model has a set of critical points which are labeled by an integer $k$.
At such points, the coefficients in \eqref{large N} behave as
\begin{align}
F_g(S) \sim (\widetilde{S}^{(k)}_c - \widetilde{S})^{(2 - \gamma) (2-2g)/2},
\end{align}
where $\widetilde{S}^{(k)}_c$ is the $k$-th critical value of the t' Hooft coupling and $\gamma = - 1/k$ is susceptibility.
Therefore, around this point, we can also expand the free energy by sending the parameters to their critical values together with the large $N$ limit
\begin{align}
\mathcal{F} = \sum_{g = 0}^\infty \kappa^{2g - 2} f_g (c),\qquad 1 - \widetilde{S} /\widetilde{S}^{(k)}_c = a^{2} c,\quad \kappa^{-1} \equiv N a^{2 - \gamma}.
\end{align}
Here $a$ is the auxiliary parameter, and $\kappa$ is kept fixed under $a \rightarrow 0$.

The free energy depends on the coefficients $A_n$ and $B_n$,
which are controlled by the recursion relations called string equations.
They are given by a set of identities
\begin{align}
	0 = \oint dz \frac{\partial}{\partial z} \left\{ \frac{z^k}{2\pi i} e^{W(z)} p_\ell(z) \tilde{p}_m(1/z) \right\}.
\end{align}
Therefore, the critical behavior of the free energy can be evaluated by solving a set of string equations at the critical point.
In particular, the cases $(k,\,\ell,\,m) = (-1,\,n,\,n-1)$ and $(k,\,\ell,\,m) = (0,\,n,\,n)$ are important:
\begin{align}	
	\oint d\mu(z) W^\prime(z) p_n(z) \tilde{p}_{n-1}(1/z) =& n (h_n - h_{n-1}),\label{streq1}\\
	\oint d\mu(z) z W^\prime(z) p_n(z) \tilde{p}_{n}(1/z) =& 0. \label{streq2}
\end{align}
For later convenience, we introduce the new variables $H_n,\, R_n$ and $G_n$ by 
\begin{align}
H_n = \sqrt{\frac{h_n}{h_{n-1}}},\qquad A_n = R_n D_n,\qquad B_n = \frac{R_n}{D_n},\qquad \frac{D_{n+1}}{D_n} = 1 + G_n.
\end{align}
In terms of these variables, \eqref{relhAB} reads
\begin{align}
	H_n^2 = 1 - R_n^2. \label{relhR}
\end{align}
Appropriate bases for \eqref{streq1} and \eqref{streq2} are respectively
\begin{align}
\mathcal{F}^{\pm}_p (H_m,G_n)  \equiv& \frac{1}{h_n}\oint d\mu(z) \left( z^{p-1} \mp \frac{1}{z^{p+1}} \right) p_n(z) \tilde{p}_{n-1}(1/z), \label{F}\\
\mathcal{G}^{\pm}_p (H_m,G_n) \equiv& \frac{1}{h_n}\oint d\mu(z) \left( z^{p} \mp \frac{1}{z^{p}} \right) p_n(z) \tilde{p}_{n}(1/z). \label{G}
\end{align}
Using these equations, \eqref{streq1} and \eqref{streq2} can be written as
\begin{align}
	\widetilde{S} \frac{n}{N} =& \frac{1 - R_n^2}{2 R_n^2} \sum_{p=1}^\infty \left( g^+_p \mathcal{F}^+_p (R,G) + g^-_p \mathcal{F}^-_p (R,G) \right),\\
	0 =& \frac{1 - R_n^2}{2 R_n^2} \sum_{p=1}^\infty\left( g^+_p \mathcal{G}^+_p (R,G) + g^-_p \mathcal{G}^-_p(R,G) \right).
\end{align}
Here we have used  \eqref{relhR} to write the functions $\mathcal{F}, \mathcal{G}$ in terms of $R_n, G_n$.

To derive the explicit form of \eqref{F} and \eqref{G}, we follow \cite{PS2} with small modification.
Let us define the normalized orthogonal polynomial
\begin{align}
	P_n(z) \equiv \frac{1}{\sqrt{h_n}}p_n(z),\qquad\widetilde{P}_n(1/z) \equiv \frac{1}{\sqrt{h_n}} \tilde{p}(1/z).
\end{align}
They are orthonormal with respect to $d\mu(z)$
\begin{align}
	\oint d\mu(z) P_n(z) \widetilde{P}_m(1/z) = \delta_{n,m}.
\end{align}
Let us introduce the operators $\hat{\ell}$ and $\hat{u}$ by
\begin{align}
	\hat{\ell} P_n(z) = n P_n(z),\qquad \hat{u} P_n(z) = P_{n+1}(z).
\end{align}
Let us write $z$ and $1/z$ in terms of $\hat{\ell}$ and $\hat{u}$.
From \eqref{rec1} and \eqref{rec2}, we obtain
\begin{align}
&\oint d\mu P_m(z) \widetilde{P}_n(1/z) z \nonumber\\
&=\left\{
\begin{array}{cc}
H_{m + 1} & n = m+1\\
-R_{m+1}  R_{m}(1 + G_m)  & n = m\\
-R_{m+1}  H_m(1 + G_m) \cdots H_{n+1}(1+G_{n+1}) R_n(1+G_n) & n < m
\end{array}
\right. ,\\
&\oint d\mu P_m(z) \widetilde{P}_n(1/z) z^{-1} \nonumber\\
&=\left\{
\begin{array}{cc}
H_{m} & n = m-1\\
-R_{m+1}  R_{m} (1 + G_m)^{-1} & n = m\\
-R_m(1+G_m)^{-1} H_{m+1} (1 + G_{m+1})^{-1} \cdots  H_{n} (1+G_n)^{-1} R_{n+1} & n > m
\end{array}
\right. . 
\end{align}
Therefore, denoting $H(n) = H_n$, ... etc, we have
\begin{align}
z =&  H(\hat{\ell}) \hat{u} - R(\hat{\ell}) (1 + G(\hat{\ell}) )\hat{u}^{-1} \frac{1}{1- H(\hat{\ell}) (1 + G(\hat{\ell})) \hat{u}^{-1} } R(\hat{\ell}) \hat{u}, \\
\frac{1}{z} =& \hat{u}^{-1} H (\hat{u}) -  \hat{u}^{-1} R(\hat{\ell}) \hat{u} \frac{1}{1 - H(\hat{\ell}) (1 + G(\hat{\ell}))^{-1} \hat{u} } \frac{R(\hat{\ell})}{1 + G(\hat{\ell})}.
\end{align}
We are interested in the planar limit where the coefficients $H_n$, $R_n$ and $G_n$ turn into the continuous function $H_n \rightarrow H(x)$, $R_n \rightarrow R(x),\, G_n \rightarrow G(x),\, n/N \rightarrow x$.
The operators $z$ and $1/z$ in this limit become respectively
\begin{align}
z =  - \frac{ 1 + G(x) - H(x) u }{ 1 - H(x)(1 + G(x)) u^{-1} }, \qquad
\frac{1}{z} = - \frac{ (1 + G(x))^{-1}  - H(x) u^{-1}}{1 - H(x) (1 + G(x))^{-1} u}.
\end{align}
Substitute these representations into \eqref{F} and \eqref{G}, we obtain the planar form of $\mathcal{F}^{\pm}_p (R, G)$ and $\mathcal{G}^{\pm}_p (R, G)$ by extracting appropriate coefficient in $u$:
\begin{align}
	\mathcal{F}^{+}_p (H,G) =& \frac{1}{H_n} \oint d\mu(z) \left( z^{p-1} - \frac{1}{z^{p+1}} \right) P_n(z) \widetilde{P}_{n-1}(1/z)\nonumber\\
	\rightarrow& \frac{1}{H} \left\{ \left. \left(- \frac{ 1 + G - H u }{ 1 - H(1 + G) u^{-1} } \right)^{p-1} \right|_{u^{-1}} - \left. \left( - \frac{ (1 + G)^{-1}  - H u^{-1}}{1 - H (1 + G)^{-1} u} \right)^{p+1} \right|_{u^{-1}} \right\} \nonumber\\
	=& (1 - H^2) \left( (1 + G)^{p} \sum_{n=0}^{p-1} (-1)^{p+n-1} \frac{ \Gamma(p + n+1) }{ p\, \Gamma(n+1) \Gamma(n+2) \Gamma(p-n-1) } H^{2n} \right. \nonumber\\
	&\qquad + \left. \frac{1}{(1 + G)^p} \sum_{n=0}^{p-1} (-1)^{p + n -1} \frac{ \Gamma(p + n + 2) }{p\, \Gamma(n+1) \Gamma(n+2) \Gamma(p-n)} H^{2n} \right),
\end{align}
\begin{align}
	\mathcal{F}^{-}_p(H,G) =& (1 - H^2) \left( (1 + G)^{p} \sum_{n=0}^{p-1} (-1)^{p+n-1} \frac{ \Gamma(p + n+1) }{ p\, \Gamma(n+1) \Gamma(n+2) \Gamma(p-n-1) } H^{2n} \right. \nonumber\\
	&\qquad - \left. \frac{1}{(1 + G)^p} \sum_{n=0}^{p-1} (-1)^{p + n -1} \frac{ \Gamma(p + n + 2) }{p\, \Gamma(n+1) \Gamma(n+2) \Gamma(p-n)} H^{2n} \right),
\end{align}
and
\begin{align}
	\mathcal{G}^{\pm}_p (H, G) =& \left\{ (1 + G)^{p} \mp \frac{1}{(1 + G)^p} \right\}  (1- H^2) \sum_{n=0}^{p-1} (-1)^{p+n} \frac{ \Gamma(p+n+1) }{ p\, \Gamma(n+1)^2 \Gamma(p-n) } H^{2n}.
\end{align}
For later convenience, let us write the planar string equations as
\begin{align}
\vec{s}(x) = \left( \frac{1-R^2}{2 R^2} \left( \sum_{p=1}^\infty \sum_{i=+,-} g^i_p \mathcal{F}^i_p (R, G) \right), \frac{1-R^2}{2R^2} \left( \sum_{p=1}^\infty \sum_{i=+,-} g^i_p \mathcal{G}^i_p (R, G) \right) \right)^t, \label{2 st eqs}
\end{align}
where  $\vec{s}(x) = (\widetilde{S} x , 0)^t$.

Let us move on to study  critical points of this model.
While we do not give the derivation \cite{PS2},
the $k$-th multicritical potential is given by
\begin{align}
	W^{(k)}(z) = -\frac{1}{2\underline{g}_s} \sum_{p=1}^k \frac{t^{(k)}_p}{p} \left( z^p + \frac{1}{z^p} \right),
\end{align}
where the critical values of the couplings at this critical point are
\begin{align}
	\left. g^+_p \right|_{\rm crit} = t^{(k)}_p = (-1)^{p+1}  \frac{ \Gamma(k) \Gamma (k+2) }{\Gamma(k-p+1) \Gamma(k+p+1)}  , \quad 1 \leq p \leq k \label{crit coupling}
\end{align}
and zero otherwise.\footnote{Here, we normalize the critical coupling $t^{(k)}_1$ to be $1$. This normalization is different from \cite{PS2}.}

The set of string equations can be expanded around $R = 0$ and $G=0$ as
\begin{align}
\vec{s}(x) =& \sum_{n=0}^\infty \sum_{r+s=n} \frac{R^r G^s}{r! s!} \nonumber\\
& \times \partial^r_R \partial^s_G 
\left( \frac{1-R^2}{2R^2} \left( \sum_{p=1}^\infty \sum_{i=+,-} g^i_p \mathcal{F}^i_p (R, G) \right), \frac{1-R^2}{2R^2} \left( \sum_{p=1}^\infty \sum_{i=+,-} g^i_p \mathcal{G}^i_p (R, G) \right) \right)^t
. \label{2 st expand}
\end{align}
By a tedious but straightforward calculation, \eqref{2 st expand} is resummed to take the following form:
\begin{align}
\vec{s}(x) = \vec{s}^{(k)}_c + a^{2-1/k} \sum_{n=1}^k \vec{a}^{(k)}_n r^{2(k-n)} g^{2n-1} + a^{2} \sum_{n=0}^k \vec{b}^{(k)}_n r^{2(k-n)} g^{2n} + \mathcal{O}(a^{2+1/k}). \label{2 critst expand}
\end{align}
Here we have set $R = a^{1/k} r, \, G = a^{1/k} g$ and introduced
\begin{align}
	\vec{s}^{(k)}_c = \left(
	\begin{array}{cc}
		\widetilde{S}^{(k)}_c\\
		0
	\end{array}
	\right) = \left(
	\begin{array}{cc}
		\dfrac{k+1}{2k}\\
		0
	\end{array}
	\right)
\end{align}
as well as
\begin{align}
\vec{a}^{(k)}_n =&\dfrac{(-1)^n}{2} \dfrac{\Gamma(k) \Gamma(k+2)}{ \Gamma(2n) \Gamma(k-n+1) \Gamma(k-n+2)} \left( 
\begin{array}{ccc}
1\\
1
\end{array}
\right),
\end{align}
and
\begin{align}
\vec{b}^{(k)}_n =& \dfrac{(-1)^{n+1}}{2} \dfrac{\Gamma(k) \Gamma(k+2)}{ \Gamma(2n+1) \Gamma(k-n+1) \Gamma(k-n+2)}  \left(
\begin{array}{ccc}
(k-n+1) + n (2n-1)\\
n (2n-1)
\end{array}
\right).
\end{align}
Eq.\eqref{2 critst expand} gets further simplified by multiplying by
\begin{align}
T = \left(
\begin{array}{cc}
1 &-1\\
0&1
\end{array}
\right),
\end{align}
as $T \vec{s}(x) = \vec{s}(x)$ and $T \vec{s}_c = \vec{s}_c$ :
\begin{align}
T \vec{a}^{(k)}_n =& \dfrac{(-1)^n}{2} \dfrac{\Gamma(k) \Gamma(k+2)}{ \Gamma(2n) \Gamma(k-n+1) \Gamma(k-n+2)}\left( 
\begin{array}{ccc}
0\\
1 
\end{array}
\right),\\
\nonumber\\
T \vec{b}^{(k)}_n =&\dfrac{(-1)^{n+1}}{2} \dfrac{\Gamma(k) \Gamma(k+2)}{ \Gamma(2n+1) \Gamma(k-n+1)^2}\left(
\begin{array}{ccc}
1\\
\dfrac{n(2n-1)}{k-n+1} 
\end{array}
\right).
\end{align}
The first component of \eqref{2 critst expand} in this form is $\mathcal{O}(a^{2})$ and the second component is $\mathcal{O}(a^{2-1/k})$.
Defining
\begin{align}
	1 - \widetilde{S} x/\widetilde{S}^{(k)}_c = a^2 t,\qquad 1 - \widetilde{S}/\widetilde{S}^{(k)}_c = a^2 c,
\end{align}
we obtain a set of planar string equations by the scaling variables $t, r$ and $g$
\begin{align}
t =&  \sum_{n=0}^{k} (-1)^{n} \dfrac{\Gamma(k+1)^2}{ \Gamma(2n+1) \Gamma(k-n+1)^2} r^{2(k-n)} g^{2n}, \label{k crit string1 in rg}\\
0 = & \sum_{n=1}^k (-1)^n \dfrac{\Gamma(k+1)^2}{ \Gamma(2n) \Gamma(k-n+1) \Gamma(k-n+2)} r^{2(k-n)} g^{2n-1}. \label{k crit string2 in rg}
\end{align}
Note that $g = 0$ is always a solution to \eqref{k crit string2 in rg}, where \eqref{k crit string1 in rg}  reduces to
\begin{align}
t = r^{2k}. \label{k crit string}
\end{align}

We can now compute the $k$-th critical behavior of the planar free energy\footnote{Note that our definition of the free energy in \eqref{large N} omits the minus sign. This explains a few minus signs in what follows.} from \eqref{planar free}
\begin{align}
	F_0 \sim&  a^{4 + 2/k} \int_{a^{-2} + c}^c dt \left\{ (t-c) t^{1/k} + \mathcal{O}(a^{2}) \right\} \nonumber\\
	\sim& -a^{4+2/k}  \frac{k^2}{(2k+1)(k+1)} c^{2 + 1/k}.
\end{align}
Then defining $\kappa^{-1} = N a^{2 + 1/k}$ which is kept fixed under the limit $N\rightarrow \infty,\,a \rightarrow 0$, we obtain the planar free energy at the $k$-th multicritical point
\begin{align}
	\mathcal{F} = -\kappa^{-2} \frac{k^2}{(2k+1)(k+1)}c^{2+1/k} + \cdots,
\end{align}
namely
\begin{align}
		f_0 = -\frac{k^2}{(2k + 1)(k+1)}c^{2 + 1/k}. \label{f_0}
\end{align}

\subsection{perturbation for critical points} \label{scalingop}
In this section, we will construct the scaling operators which perturb the  $k$-th critical planar string equations \eqref{k crit string1 in rg} and \eqref{k crit string2 in rg}.
Let us first review
how to construct the scaling operators which preserve the $z \rightarrow 1/z$ invariance \cite{PS2}.
We will call them even type scaling operators.
We also construct the odd and logarithmic type scaling operators which are odd under $z \rightarrow 1/z$.

\subsubsection{even type perturbation}
Let us consider the following perturbation of the $k$-th critical potential
\begin{align}
	W^{(k)}(z) \rightarrow W^{(k)}(z) + m^{+ B}_\ell \sigma^{+B}_\ell(z),\qquad \sigma^{+B}_\ell(z) = W^{(\ell)} (z) ,
\end{align}
where $m^{+B}_\ell$ is ``bare" coupling and $\sigma^{+B}_\ell(z)$ is ``bare" scaling operator.
It is clear that perturbed potential is invariant under $z \rightarrow 1/z$.
This perturbation changes the string equation \eqref{2 critst expand} into
\begin{align}
	\vec{s}(x) =& \vec{s}^{(k)}_c + a^{2-1/k} \sum_{n=1}^k T \vec{a}^{(k)}_n r^{2(k-n)} g^{2n-1} + a^{2} \sum_{n=0}^k T \vec{b}^{(k)}_n r^{2(k-n)} g^{2n} \nonumber\\
	&+ m^{+B}_\ell \left( \vec{s}^{(\ell)}_c + a^{(2\ell-1)/k} \sum_{n=1}^\ell T \vec{a}^{(\ell)}_n r^{2(\ell-n)} g^{2n-1} + a^{2 \ell/k} \sum_{n=0}^\ell T \vec{b}^{(\ell)}_n r^{2(\ell-n)} g^{2n} \right). \label{even pert st eqs}
\end{align}
Since the critical value of $\widetilde{S}^{(k)}_c$ shifts by
\begin{align}
	\widetilde{S}^{(k)}_c \rightarrow \widetilde{S}^{(k)\prime}_c = \widetilde{S}^{(k)}_c + m^{+B}_\ell \widetilde{S}^{(\ell)}_c,
\end{align}
we redefine the scaling variable for $\widetilde{S}$ by
\begin{align}
	1 - \widetilde{S} x / \widetilde{S}^{(k) \prime}_c = a^2 t^\prime.
\end{align}
In order for the first component of \eqref{even pert st eqs} to be oder $a^2$, we should scale $m^{+B}_\ell$ by
\begin{align}
	m^{+B}_\ell = a^{2(1 - \ell/k)} \frac{\widetilde{S}^{(k)}_c}{\widetilde{S}^{(\ell)}_c} m^+_\ell.
\end{align}
For $\ell = 1,\cdots, k-1$, the critical value $\widetilde{S}^{(k)\prime}_c$ goes to $\widetilde{S}^{(k)}_c$ in the $a \rightarrow 0$ limit.
It means that the critical value of $\widetilde{S}^{(k)}_c$ does not change at the $k$-th multicritical point.
For $\ell = k$, the perturbation changes only the critical values of $\widetilde{S}^{(k)}$.
For $\ell > k $, we must set $m^+_\ell = 0$ to make $\widetilde{S}^{(k) \prime}_c$ into finite.
We should therefore consider the $\ell = 1,\cdots,k-1$ cases,
where the perturbed string equations read
\begin{align}
	 t=& \left(  \sum_{n=0}^{k} (-1)^{n} \dfrac{\Gamma(k+1)^2}{ \Gamma(2n+1) \Gamma(k-n+1)^2} r^{2(k-n)} g^{2n}\right) \nonumber\\
	 &\quad +  m^+_\ell \left( \sum_{n=0}^{\ell} (-1)^{n} \dfrac{\Gamma(\ell+1)^2}{ \Gamma(2n+1) \Gamma(\ell-n+1)^2} r^{2(\ell-n)} g^{2n} \right)  ,\label{even pert crit string1}\\
	0 =&  \left( \sum_{n=1}^k (-1)^n \dfrac{\Gamma(k+1)^2}{ \Gamma(2n) \Gamma(k-n+1) \Gamma(k-n+2)} r^{2(k-n)} g^{2n-1} \right) \nonumber\\
	&\quad +  m^+_\ell  \left( \sum_{n=1}^\ell (-1)^n \dfrac{\Gamma(\ell+1)^2}{ \Gamma(2n) \Gamma(\ell-n+1) \Gamma(\ell-n+2)} r^{2(\ell-n)} g^{2n-1} \right) . \label{even pert crit string2}
\end{align}
Note that \eqref{even pert crit string2} has a solution $g = 0$ for $\ell = 1,\cdots,k-1$ and \eqref{even pert crit string1} becomes
\begin{align}
	t = m^+_\ell r^{2\ell} + r^{2k}.
\end{align}

\subsubsection{odd type perturbation}
Let us now consider the following perturbation
\begin{align}
W^{(k)}(z) \rightarrow W^{(k)}(z) + m^{-B}_\ell \sigma^{-B}_\ell(z),
\end{align}
where $1 \leq \ell \leq k -1$ and
\begin{align}
\sigma^{-B}_\ell(z) = -\frac{1}{2 \underline{g}_s} \sum_{p=1}^{\ell} \bar{t}^{(\ell)}_p \left(  z^p - \frac{1}{z^p}\right),\qquad \bar{t}^{(\ell)}_p = p \, t^{(\ell)}_p.
\end{align}
Here $t^{(\ell)}_p$ is given by \eqref{crit coupling}.
The perturbed string equation is
\begin{align}
\vec{s}(x) =& \left( \frac{1-R^2}{2 R^2} \left( \sum_{p=1}^k  t^{(k)}_p \mathcal{F}^+_p (R, G) \right), \frac{1-R^2}{2R^2} \left( \sum_{p=1}^k  t^{(k)}_p \mathcal{G}^+_p (R, G) \right) \right)^t \nonumber\\
&+ m^{-B}_\ell \left( \frac{1-R^2}{2 R^2} \left( \sum_{p=1}^\ell  \bar{t}^{(\ell)}_p \mathcal{F}^-_p (R, G) \right), \frac{1-R^2}{2R^2} \left( \sum_{p=1}^\ell  \bar{t}^{(\ell)}_p \mathcal{G}^-_p (R, G) \right) \right)^t.
\end{align}
The first line has been already computed in \eqref{2 critst expand}.
The second line can be resummed to give
\begin{align}
m^{-B}_\ell a^{(2\ell - 2)/k} \sum_{n=1}^{\ell} \vec{c}_n r^{2(\ell-n)} g^{2n -2} + m^{-B}_\ell a^{(2\ell-1)/k} \sum_{n=1}^{\ell} \vec{d}_n r^{2(\ell-n)} g^{2n-1} + \mathcal{O}(a^{2\ell/k}),
\end{align}
where
\begin{align}
\vec{c}_n =& \frac{(-1)^{n}}{2} \dfrac{ \Gamma(\ell) \Gamma(\ell+2) }{ \Gamma(2n-1) \Gamma(\ell-n+1) \Gamma(\ell-n+2)  }  \left(
\begin{array}{ccc}
1\\
1
\end{array}
\right),\\
\nonumber\\
\vec{d}_n =& \frac{(-1)^{n+1}}{2} \dfrac{  \Gamma(\ell) \Gamma(\ell+2) }{ \Gamma(2n) \Gamma(\ell-n+1) \Gamma(\ell-n+2)  } \left(
\begin{array}{ccc}
 1 +\ell- n + (n-1) (2n -1) \\
(n-1)(2n-1) 
\end{array}
\right).
\end{align}
These can be also transformed into a simpler form by the multiplication by $T$:
\begin{align}
T \vec{c}_n =&  \frac{(-1)^{n}}{2} \dfrac{ \Gamma(\ell) \Gamma(\ell+2) }{\Gamma(2n-1) \Gamma(\ell-n+1) \Gamma(\ell-n+2)  } \left(
\begin{array}{cc}
0\\
1
\end{array}
\right),\\
T \vec{d}_n =& \frac{(-1)^{n}}{2} \frac{\Gamma(\ell) \Gamma(\ell+2)}{\Gamma(2n) \Gamma(\ell-n+1)^2} \left(
\begin{array}{cc}
1\\
\dfrac{(n-1)(2n-1)}{\ell-n+1}
\end{array}
\right).
\end{align}
Thus, this perturbation shifts the first string equation at $\mathcal{O}(a^{(2\ell-1)/k})$ and the second string equation at $\mathcal{O}(a^{(2\ell-2)/k})$.
In order to make the first string equation $\mathcal{O}(a^2)$, we should rescale the coupling $m^{-B}_\ell$ as
\begin{align}
m^{-B}_\ell = a^{2 - (2\ell-1)/k} \frac{\widetilde{S}^{(k)}_c}{\widetilde{S}^{(\ell)}_c} m^-_\ell.
\end{align}
The resulting perturbed string equations are
\begin{align}
t =&  \sum_{n=0}^{k} (-1)^{n} \dfrac{\Gamma(k+1)^2}{ \Gamma(2n+1) \Gamma(k-n+1)^2} r^{2(k-n)} g^{2n} \nonumber\\
&+ m^-_\ell \sum_{n=1}^{\ell} (-1)^{n+1} \frac{\Gamma(\ell + 1)^2}{\Gamma(2n) \Gamma(\ell-n+1)^2} r^{2(\ell-n)}g^{2n-1} , \label{odd pert k crit string1 in rg}\\
0 = & \sum_{n=1}^k (-1)^n \dfrac{\Gamma(k+1)^2}{ \Gamma(2n) \Gamma(k-n+1) \Gamma(k-n+2)} r^{2(k-n)} g^{2n-1} \nonumber\\
&+ m^-_\ell \sum_{n=1}^{\ell} (-1)^{n} \frac{\Gamma(\ell+1)^2}{ \Gamma( 2n -1) \Gamma(\ell-n+1) \Gamma(\ell-n+2) } r^{2(\ell-n)}g^{2n-2}. \label{odd pert k crit string2 in rg}
\end{align}

\subsubsection{logarithmic type perturbation}
Let us finally consider the following perturbation
\begin{align}
	W^{(k)}(z) \rightarrow W^{(k)}(z) + m^{B}_{\rm log} \sigma^B_0(z),\qquad \sigma^B_{\rm log}(z) = -\frac{1}{2 \underline{g}_s} \log(z). \label{log pert}
\end{align}
Because
\begin{align}
	\frac{1}{h_n}\oint d\mu(z) \frac{1}{z} p_n(z) \tilde{p}_{n-1}(1/z) = 	\frac{1}{h_n}\oint d\mu(z)  p_n(z) \tilde{p}_{n}(1/z) = 1,
\end{align}
the net effect of this perturbation is just to shift r.h.s. of \eqref{2 critst expand} by 
\begin{align}
	\frac{1-R^2}{R^2} (m^B_{\rm log}, m^B_{\rm log})^t.
\end{align}
The $k$-th critical string equations perturbed by \eqref{log pert} are
\begin{align}
	\vec{s}(x) = \vec{s}_c + \left(a^{-2/k} r^{-2} - 1\right) (0, \,m^B_0)^t + a^{2-1/k} \sum_{n=1}^k T \vec{a}_n r^{2(k-n)} g^{2n-1} + a^{2} \sum_{n=0}^k T \vec{b}_n r^{2(k-n)} g^{2n}. \label{log pert critst}
\end{align}
This perturbation affects the second string equation only.
For the second component of \eqref{log pert critst} to be $\mathcal{O}(a^{2-1/k})$, we should scale $m^B_0$ as
\begin{align}
	m^B_{\rm log} = a^{2 + 1/k} m_{\rm log}.
\end{align}
The perturbed string equations of the logarithmic type are
\begin{align}
	t =&  \sum_{n=0}^{k} (-1)^{n} \dfrac{\Gamma(k+1)^2}{ \Gamma(2n+1) \Gamma(k-n+1)^2} r^{2(k-n)} g^{2n}, \label{log perturbed kst1}\\
	0 =& m_{\rm log} + \sum_{n=1}^k (-1)^n \dfrac{\Gamma(k+1)^2}{ \Gamma(2n) \Gamma(k-n+1) \Gamma(k-n+2)} r^{2(k-n + 1)} g^{2n-1} \label{log perturbed kst2}
\end{align}

\subsection{scaling behavior and the scaling dimension} \label{scalingops}

\subsubsection{scaling operators}
In section \ref{scalingop}, we derived the perturbed string equations obtained by the various perturbations.
The vevs of the scaling operators at the $k$-th multicritical point are given by the derivative of free energy with respect to the scaled perturbation parameter:
\begin{align}
	\left< \sigma^{\pm}_\ell \right> = \frac{\partial \mathcal{F}}{\partial m^\pm_\ell},\qquad 
		\left< \sigma_{\rm log} \right> = \frac{\partial \mathcal{F}}{\partial m_{\rm log}}.
\end{align}
Since they can be also expanded in $\kappa$ as is the free energy, their leading parts are written as
\begin{align}
	\left< \sigma^\pm_{\ell,0} \right> = \frac{\partial f_0}{\partial m^\pm_\ell},\qquad \left< \sigma_{{\rm log},0} \right> = \frac{\partial f_0}{\partial m_{\rm log}}.
\end{align}
We can evaluate $\left< \sigma^\pm_{\ell,0} \right> $ and $\left< \sigma_{{\rm log},0} \right>$ by solving the perturbed string equations.

Let us consider the even type scaling operators which are studied in \cite{PS2}:
\begin{align}
\left< \sigma^+_{\ell, 0} \right> = \frac{\partial f_0}{\partial m^+_\ell} \sim  \int^c dt\, (t-c)\,  \frac{ \partial r(t ; \{m^+_\ell\})^{2} }{\partial m^+_\ell},
\end{align}
where $r(t ; \{m^+_\ell\})$ is the solution of 
\begin{align}
	t = \sum_{\ell=1}^{k-1} m^+_\ell r^{2\ell} + r^{2k},
\end{align}
which can be obtained in the same way as in \cite{GM}:
\begin{align}
r(t ; \{m^+_\ell\})^2 = t^{1/k} + \sum_{p=1}^\infty \frac{(-1)^p}{k \, p!} \left( \frac{\partial}{\partial t} \right)^{p-1} \left\{ \left( \sum_{\ell=1}^{k-1} m^+_\ell t^{\ell/k} \right)^p t^{-1 + 1/k} \right\}.
\end{align}
We obtain
\begin{align}
\left<\sigma^+_{\ell, 0} \right> \sim  \frac{k}{(k + \ell + 1)(\ell + 1)} c^{1 + (\ell + 1)/k} + \mathcal{O}(m^+_\ell). \label{sigma+}
\end{align}

To evaluate the odd type scaling operators, we solve the set of equations \eqref{odd pert k crit string1 in rg} and \eqref{odd pert k crit string2 in rg}.
We assume that $m^-_\ell$ is small and $g$ can be expanded in $m^-_\ell$ as
\begin{align}
g(r ;m^-_\ell) = \sum_{n=1}^\infty g_n(r) (m^-_\ell)^{2n-1}. \label{g sol in odd}
\end{align} 
We choose the solution of \eqref{odd pert k crit string2 in rg} to satisfy $g(r ; m^-_\ell = 0) = 0$ so that it reduces to \eqref{k crit string} when there is no deformation.
Similarly, we assume that $r$ has the following expansion
\begin{align}
r(t ; m^-_\ell)^2 = t^{1/k} \left( 1 + \sum_{n=1}^\infty r_n(t) (m^-_\ell)^{2n} \right)
\end{align}
such that $r = t^{1/(2k)}$ at $m^-_\ell = 0$.
Substituting \eqref{g sol in odd} into \eqref{odd pert k crit string1 in rg}, and solving oder by oder, we obtain
\begin{align}
r(t ; m^-_\ell)^2 = t^{1/k} + \frac{\ell^2}{2k^2} ( k + 2\ell) t^{-2 + 2\ell/k} (m^-_\ell)^2 + \mathcal{O}((m^-_\ell)^4),
\end{align}
and
\begin{align}
\left<\sigma^-_{\ell,0} \right> = \frac{\partial f_0}{\partial m^-_\ell} =& \frac{\ell^2}{k^2} (k+ 2\ell) m^-_\ell \int^c_{a^{-2}+c} dt (t-c) t^{-2 + 2\ell/k} + \mathcal{O}((m^-_\ell)^3) \nonumber\\
=& \frac{\ell (k+2\ell)}{2 (k-2\ell)} c^{2\ell/k} m^-_\ell + \mathcal{O}((m^-_\ell)^3). \label{sigma-}
\end{align}

Similarly, for the logarithmic type scaling operator, we obtain from the solution of \eqref{log perturbed kst1} and \eqref{log perturbed kst2}
\begin{align}
	r(m_{\rm log})^2 = t^{1/k} + \frac{1}{2 t^2} (m_{\rm log})^2 + \mathcal{O}((m_{\rm log})^4),
\end{align}
and
\begin{align}
	\left<\sigma_{{\rm log},0} \right> = \frac{\partial f_0}{\partial m_{\rm log}} =& m_{\rm log} \int^c_{a^{-2} + c} dt (t-c) t^{-2} + \mathcal{O}((m_{\rm log})^3) \nonumber\\
	\sim& m_{\rm log} \left( \log c +1  \right) + \mathcal{O}((m_{\rm \log})^3). \label{sigmalog}
\end{align}

\subsubsection{scaling dimensions}

We search for the possibility that the free energy at the $k$-th multicritical point is that of the AD theory of some type.
Then, we define the scaling dimension of the sphere contribution of the free energy $f_0$ to be $\left[ f_0 \right] = 2$, since it should be identified the Seiberg-Witten prepotential of corresponding AD theory.
The scaling dimensions of the other variables are easily determined from the critical behavior of the free energy \eqref{f_0} and the vevs of the scaling operators \eqref{sigma+}, \eqref{sigma-} and \eqref{sigmalog}:
\begin{align}
	\left[ c \right] = \frac{2k}{2k+1},
\end{align}
\begin{align}
	\left[\sigma^+_\ell \right] = \frac{2k + 2\ell + 2}{2k + 1},\quad \left[\sigma^-_{\ell,0} \right] = \frac{2k + 2\ell + 1}{2k + 1},\quad \left[ \sigma_{{\rm log},0} \right] = 1,\quad \ell = 1,\ldots,k-1,	 
\end{align}
and
\begin{align}
	\left[ m^+_\ell \right] = \frac{2k - 2 \ell}{2k + 1}, \quad	[m^-_\ell] = \frac{2k-2\ell+1}{2k+1}, \quad \left[ m_{\rm log} \right]	= 1, \quad \ell = 1,\ldots,k-1.
\end{align}

\section{Comparison with Argyres-Douglas theory}

Let us compare the scaling operators which we constructed in section \ref{scalingops} with those of the Coulomb branch operators in the AD theory.
In this way, we can easily see the relationship between the $k$-th multicritical point and the AD theory.
In fact, in the $(A_1, A_{4k-1})$ theory, the scaling dimension for the Coulomb branch operators $u_{2k+1+i}$ and their coupling constant $m_i \equiv u_{2k + 1 - i}$ are respectively\footnote{Here, the Seiberg-Witten curve of the $(A_1, A_{4k-1})$ AD theory is given by $x^2 = z^{4k} + u_2 z^{4k-2} + \cdots + u_{4k}$.}
\begin{align}
[u_{2k+1+i}] = \frac{2k+1+i}{2k+1},\qquad [m_i] = \frac{2k+1-i}{2k+1},\qquad i = 1,\ldots,2k-1.
\end{align}
We obtain a dictionary by setting $i = 2\ell$ and $i = 2 \ell + 1$
\begin{align}
[\sigma^-_{\ell,0}] &=[u_{2k+1+2\ell}], & [m^-_{\ell}] &= [m_{2\ell}]  ,\qquad \ell = 1,\ldots,k-1, \\
[\sigma^+_{\ell,0}] &=[ u_{2k + 2 +2\ell} ]  ,& [m^+_\ell] &= [m_{2\ell+1}], \qquad \ell = 0,\cdots,k-1,
\end{align}
and
\begin{align}
[m_{\rm log}] = [u_{2k+1}].
\end{align}
Here we introduced
\begin{align}
\sigma^+_{0,0} \equiv \frac{\partial f_0}{\partial c},\qquad m^+_0 \equiv c.
\end{align}

The perturbations from the $k$-th multicritical point can capture all Coulomb branch operators which are contained in the $(A_1, A_{4k-1})$ theory.
We conclude that the $k$-th multicritical point at the even potential of the one unitary matrix model corresponds to the $(A_1, A_{4k-1})$ theory.

\section*{Acknowledgment}
We are indebted to Takeshi Oota for sharing understanding on various aspects of this subject.
We thank Kazumasa Okabayashi for helpful discussion.
The work of HI is supported in part by JSPS KAKENHI Grant Number 19K03828
and by the Osaka City University (OCU) Strategic Research Grant 2020 for priority area (OCU-SRG2019\_TPR01).

\end{document}